\documentclass[aps,prl,twocolumn,superscriptaddress,showpacs]{revtex4-1}

\usepackage{amssymb} 
\usepackage{graphicx}
\usepackage{color}
\usepackage[integrals]{wasysym}
\usepackage{amsmath}
\usepackage{braket}
\usepackage{soul}

\begin{document}


\title{Role of Ce 4$f$ hybridization in the origin of magnetism in nanoceria}

\author{V. K. Paidi}
\affiliation{Department of Physics and Astronomy, University of Manitoba, Winnipeg, Manitoba, R3T 2N2, Canada}

\author{D. L. Brewe}
\affiliation{Advanced Photon Source, Argonne National Laboratory, Argonne, Illinois 60439, USA}

\author{J. W. Freeland}
\affiliation{Advanced Photon Source, Argonne National Laboratory, Argonne, Illinois 60439, USA}

\author{C. A. Roberts}
\affiliation{Toyota Motor Engineering and Manufacturing North America Inc., 1555 Woodridge Avenue, Ann Arbor, Michigan 48105, USA}

\author{J. van Lierop}
\affiliation{Department of Physics and Astronomy, University of Manitoba, Winnipeg, Manitoba, R3T 2N2, Canada}
\email{paidivk@myumanitoba.ca,Johan.van.Lierop@umanitoba.ca}


\begin{abstract}
Nanoscale CeO$_2$ (nanoceria) is a prototypical system that presents d$^0$ ferromagnetism.  Using a combination of x-ray absorption spectroscopy, x-ray magnetic circular dichroism and modelling, we show that nano-structure, defects and disorder, and non-stoichiometry create magnetically polarized Ce 4$f$ and O 2$p$ hybridized states captured by the vacancy orbitals ($V_{orb}$) that are vital to ferromagnetism. Further, we demonstrate that foreign ions (Fe and Co) enhance the moment at Ce 4$f$ sites while the number of $V_{orb}$ is unchanged, pointing clearly to the mechanism of orbital hybridization being key missing ingredient to understanding the unexpected ferromagnetism in many nanoscale dilute magnetic oxides and semiconductors.

\end{abstract}


\maketitle


Defects, disorder, and non-stoichiometry are considered to be the key ingredients for d$^0$~magnetism in nanoscale wide band-gap oxides. d$^0$~magnetism has drawn significant interest as reflected by the many reproducible experimental observations of unexpected ferromagnetism in bulk-nonmagnetic oxides such as CeO$_2$, ZnO, HfO$_2$, Al$_2$O$_3$, In$_2$O$_3$, SnO$_2$ and many dilute magnetic oxides\cite{Sundaresan.2006,Straumal.2009,Chaboy.2010,Esmaeily.2018,Coey.2005,Coey.2008,dietl.2010,Green.2015}.  In general these materials have been quite puzzling due to the challenge of identifying the exact origin of the magnetism and distinguishing its spin and orbital character. It has been shown that nanoscale CeO$_2$ (nanoceria) is the prototypical system that has extensive spontaneous ferromagnetism with no magnetic cations\cite{Ackland.2018}. The physics of this magnetism has been enigmatic.  At first, the magnetism was attributed to most obvious candidate, exchange interactions between localized electron spin moments resulting from the oxygen vacancies \cite{Sundaresan.2006}; first-principles calculations revealed that the vacancies (especially those at the surface) can induce magnetic moments in nanoceria\cite{Ge.2008,Han.2009}. Later, the ferromagnetism was attributed to only sub-20~nm nanoceria with no obvious dependence on oxygen vacancies\cite{Liu.2008}. Others reported that mixed valence Ce$^{3+}$/Ce$^{4+}$ pairs on the surface were responsible\cite{Li.2009}. Recently, a model based on a giant orbital paramagnetism phenomenon\cite{Coey.2016} that occurs in a mesoscopic quasi-two-dimensional configuration with dilute magnetically active sites has been proposed. Despite $d^0$ behavior in nanoceria being widely reproducible\cite{Paunovic.2012,Sundaresan.2006,Liu.2008,Ge.2008,Ackland.2011,Chen.2012,Lee.2014,Coey.2016,Ackland.2018}, an understanding of the physics behind the nanomagnetism with the three key ingredients is still lacking.

In this work, we focus on the fundamental problem related to identifying the origin of the magnetism in nanoceria and ascertaining the mechanisms that affect the magnetic properties.  We use local probes of the electronic structure and magnetism (e.g. x-ray absorption spectroscopy and x-ray magnetic circular dichroism, M\"ossbauer spectroscopy) combined with conventional magnetometry to provide insights into the underlying physics. Although there are earlier reports on the element specific magnetism of nanoceria\cite{Chen.2012,Lee.2014,Peng.2014}, because of the weak XMCD signal explicit evidence of the spin and orbital contributions to the magnetic moments of the Ce 4$f$ states are still lacking. Using electronic structure, surface and bulk magnetism we unambiguously demonstrate that vacancy orbitals ($V_{orb}$), hybridization, spin and orbital angular momentum are fundamental to explain long range ferromagnetic order. Additionally, we have identified that foreign ions (Fe and Co) on nanoceria enhances the ferromagnetic moment at the Ce 4$f$ sites, and a microscopic mechanism is proposed to explain the origin of magnetism in nanoscale oxide semiconductors. 

Nanoceria\cite{Nanotex}, and Fe and Co decorated nanoceria were prepared as described in Ref.\cite{Roberts.2015,Peck.2017}.  The surface densities (chosen for no secondary phase formation) were 1.11 Fe/nm$^2$ and 3.57 Co/nm$^2$\cite{Roberts.2015}.  X-ray diffraction (XRD) pattern refinements yielded identical CeO$_2$ cubic Fm$\bar3$m structures for all systems (see SM). Transmission electron micrographs (TEM) and high-angle annular dark-field images were consistent with the XRD analysis.  Crystallite sizes were of the order of 20~nm in diameter and lattice constants were $5.411\pm0.001$~\AA.  XRD, TEM and M\"{o}ssbauer spectroscopy (see SM) results confirmed that no secondary phases (e.g. metal oxides) were present, as do the hard and soft x-ray absorption measurements discussed below. To identify the overall magnetism $M(\mu_0 H)$ measurements were performed. $M(\mu_0H)$ of nanoceria shows a coercivity of $\sim$50~mT and saturation magnetization ($M_s$) of $\sim$4~Am$^{-1}$. Co and Fe decorated nanoceria $M_s$'s were $\sim$4~Am$^{-1}$ and 7~Am$^{-1}$, respectively, in agreement with many reports in the literatures (see SM, Ref.~\cite{Ackland.2018} and references therein).

\begin{figure}

\begin{center}		
\includegraphics[scale=0.45]{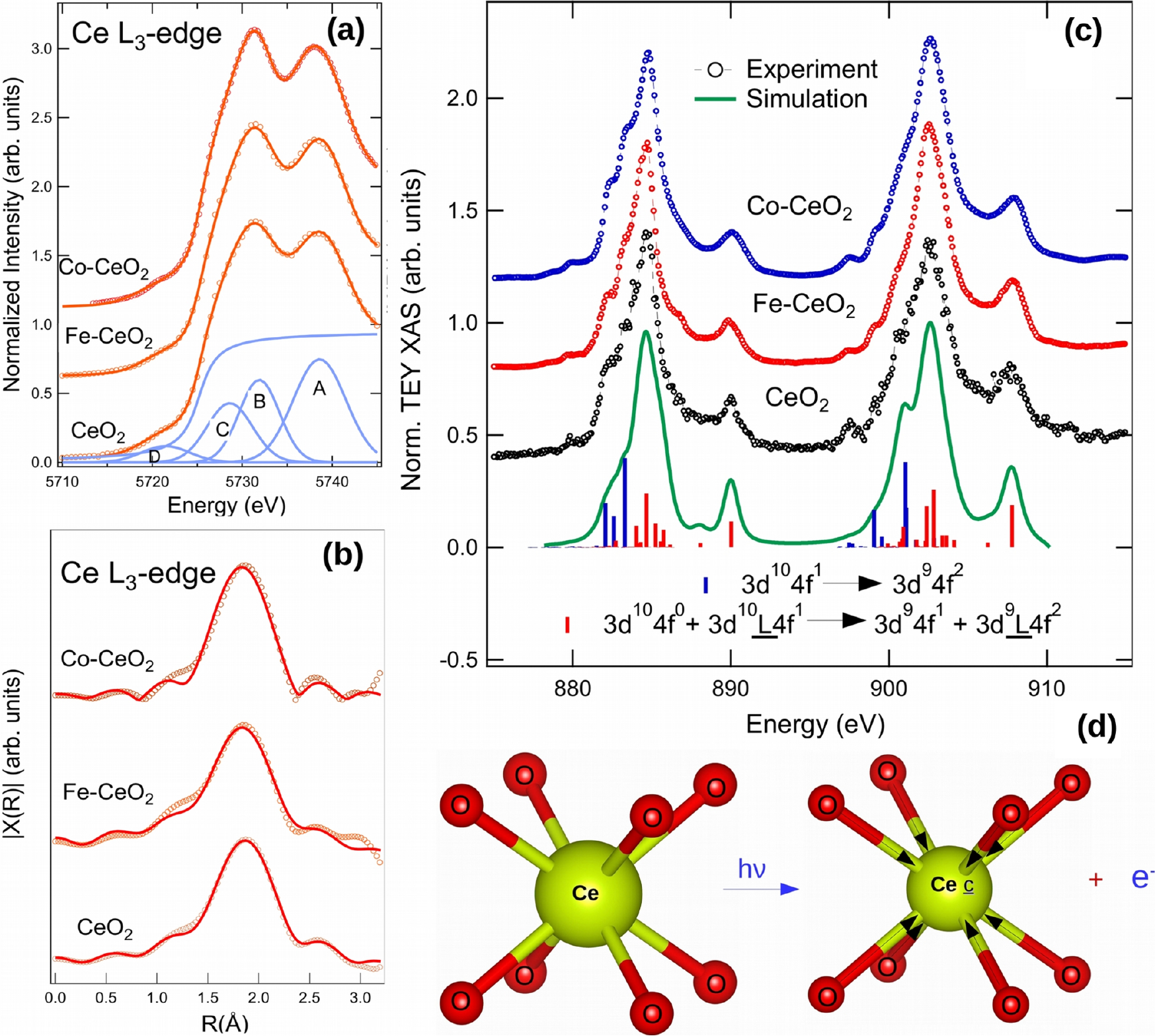}
\caption{(a) The normalized XANES spectra were fitted with Gaussian functions. To exclude the effects of the edge jump from fits an arctan function was included, as shown. (b) Fourier transforms represent raw data without correcting for phase shifts. Theoretical fits are the solid lines.   
(c) Ce $M_{4,5}$ edge XAS data and the simulation. Charge transfer effects with 4$f^0$+4$f^1\underline{L}$ ground and 4$f^1$+4$f^2\underline{L}$ final states are included in order to match the experimental spectra as discussed in the text. (d) Representation of charge transfer effect between O 2$p$ ligand and Ce 4$f$ are shown; $\underline{c}$ is core hole on Ce. \label{fig:XAFS_XAS_Ce}} 
\end{center}

\end{figure}

Because the electronic and magnetic properties of Ce ions depend strongly on the localized and delocalized 4$f$ electron states, x-ray absorption near edge structure (XANES) experiments were performed to identify and quantify the oxidation state of Ce ions in nanoceria, Fe-decorated nanoceria (Fe-CeO$_2$) and Co-decorated nanoceria (Co-CeO$_2$). As shown in figure \ref{fig:XAFS_XAS_Ce}a, XANES spectra exhibit a doublet due to the interaction between the 4$f$ orbitals of the Ce atoms and 2$p$ orbitals of oxygen ligands. The peculiar doublet consists of four observed peaks\cite{Dexpert.1987,Kaindl.1988,Soldatov.1994}.  Component A is assigned to the transition from the Ce 2$p$ shell to 5$d$ shell (final state 2$\underline{p}$4$f^0$5$d^1$ with no $f$ electrons) while component B is assigned to the excitation from the 2$p$ shell to 5$d$ shell along with an electron being excited from the O 2$p$ shell to the Ce 4$f$ shell, thus leaving a hole in the valence band (final state 2$\underline{p}$4$f^1$5$d^1$v;~v is the hole). Component C is assigned to Ce$^{3+}$, and component D is assigned to the 2$p_\frac{3}{2}$ $\rightarrow$ 4$f$ quadrupole transition that is a consequence of 5$d$ admixtures to the 4$f$ states\cite{Sham.2005}. The concentrations of Ce$^{3+}$ from spectral weighting were estimated to be $20\pm2$\%. In nanoceria each Ce atom ([Xe]4$f^1$5$d^1$6$s^2$) can donate four electrons to bonding orbitals with two O (1$s^2$2$s^2$2$p^4$) atoms. When an oxygen vacancy is formed, the two electrons previously occupying $p$ orbitals of the O atom are free to distribute. The localized electrons around Ce atoms changes the oxidation state from Ce$^{4+}$ to Ce$^{3+}$. The constant Ce$^{3+}$(4$f^1$) is as expected since the Fe and Co ions are surface decorating the nanoceria (i.e. Fe and Co ions distributed randomly on the surface of the nanoceria crystallites, bonding covalently through available O ions, as shown experimentally in Refereces \cite{Roberts.2015,Peck.2017}).

In order to gain insights into the local environment around Ce ions, we examined the extended x-ray absorption fine structure (EXAFS). Fourier transformed data and the corresponding EXAFS oscillations are shown in figure \ref{fig:XAFS_XAS_Ce}b. Spectral fits identify that the bond distances of first shell Ce -- O systems are of 2.31 $\pm$ 0.04 ${\AA}$. The coordination number (see SM for details) and structural disorder around Ce (identified by Debye-Waller factors) increases in Fe-CeO$_2$ and Co-CeO$_2$. The Ce $L_3$ edge XAS results show that for all systems, the Ce sites exist between Ce$^{3+}$ (4$f^1\underline{v}$) and Ce$^{4+}$ (4$f^0$) character, with a hole ($\underline{v}$) in the O~2$p$ valence band.

To describe the $f$ electrons, their occupancy, and electronic structure at the Ce sites, we used the Ce $M_{4,5}$ edge XAS (probes directly the valence 4$f$ states by exciting electrons from 3$d$ core orbitals, and gives insights to the ground state) in combination with atomic multiplet calculations based on a simplified Anderson impurity model\cite{DeGroot.2005,Thole.1985}. The $M_{4,5}$ edge XAS spectra (Fig. \ref{fig:XAFS_XAS_Ce}c) of nanoceria consists of main peaks at 884.6 and 902.4 eV and additional weaker satellite peaks at 889.8 and 908.0 eV. The energy splitting between Ce $M_{4,5}$ edges is due to the spin-orbit coupling with the 3$d_{\frac{5}{2}}$ and 3$d_{\frac{3}{2}}$ core-holes. The primary features of the Ce $M_{4,5}$ edge XAS spectra originate from electric-dipole allowed transitions from 3$d^{10}$4$f^n$ $\rightarrow$ 3$d^9$4$f^{n+1}$\cite{Thole.1985}. For nanoceria, experimental spectra are simulated including Coulomb, exchange, and spin-orbit interactions by considering only 3$d^{10}$4$f^0$ $\rightarrow$ 3$d^{9}$4$f^1$ and 3$d^{10}$4$f^1$ $\rightarrow$ 3$d^{9}$4$f^2$ configurations. Results indicated that if we assumed only oxygen vacancies and the ground states were due to 4$f^0$ and 4$f^1$ atomic-like multiplets, the experimental spectra could not be modelled successfully (see SM). In order to understand the Ce $M_{4,5}$ edge XAS spectra, especially the origin of the higher energy satellites, we focused on the ligand hole contribution to the 3$d^{10}$ 4$f^0$ ground state (from charge fluctuations in initial and final states due to hole on oxygen ligand). A schematic representation of a cluster consisting of a Ce ion surrounded by eight O ions is shown in Fig. \ref{fig:XAFS_XAS_Ce}d. Because of the strong Ce 4$f$ -- O 2$p$ hybridization, the initial state of the transition is described by 3$d^{10}$4$f^0$ + 3$d^{10}$\underline{L}4$f^1$ and the final state by 3$d^{9}$4$f^1$ + 3$d^{9}$\underline{L}4$f^2$ (where \underline{L} describes a hole in the O 2$p$ band\cite{Butorin.1996}). The two configurations in the final state form bonding (3$d^9$4$f^1$) and antibonding (3$d^9\underline{L}$4$f^1$) orbital combinations. Four additional terms $\Delta E_{gs}$, $T_{gs}$, $\Delta E_{fs}$, and $T_{fs}$ are defined to describe the relative energies and interactions of these initial and final states\cite{Cowan.1981,Thole.1985}. Here $\Delta E_{gs} = E(3d^{10}\underline{L}4f^1) - E(3d^{10}4f^0)$ is the charge transfer energy between two ground states, and $T_{gs} = \braket{(3d^{10}\underline{L}4f^1)|H|(3d^{10}4f^0)}$ is the effective hopping-integral connecting the two ground state configurations. Similarly $\Delta E_{fs} = E(3d^9\underline{L}4f^2)-E(3d^94f^1)$ and $T_{fs} = \braket{(3d^9\underline{L}4f^2)|H|(3d^94f^1)}$ are charge transfer and hopping integrals of the final state (see SM). Our simulation was modelled with 77\% 3$d^{10}$4$f^0$ and 23\% 3$d^{10}\underline{L}4f^1$ ground state configuration and $\Delta E_{gs} = 2.0~$eV and $T_{gs} = 0.77~$eV. The $\Delta E_{fs}$ is defined as the sum of $\Delta E_{gs} + U_{ff} - U_{fd}$, where $U_{ff}$ represent the Coulomb repulsion and $U_{fd}$ the core-valence repulsion integrals. Our simulation agrees best with the experimental data with $\Delta E_{fs} = -2.5$ eV. For a purely Ce$^{4+}$ based system the Ce $M_{4,5}$ edge $\Delta E_{fs} = -1.5$ eV\cite{Loble.2015}. In lanthanides it is expected that $U_{ff} > U_{fd}$ due to the smaller orbital radius\cite{DeGroot.2005}. However, in nanoceria, $U_{fd} > U_{ff}$ indicates that the charge transfer energy is reduced due to covalent Ce 4$f$ -- O 2$p$ states in this mixed valency system. Earlier, on the basis of band-structure calculations it was shown that ceria is less ionic\cite{Koelling.1983}.

\begin{figure}[t!]

\begin{center}		
\includegraphics[scale=0.48]{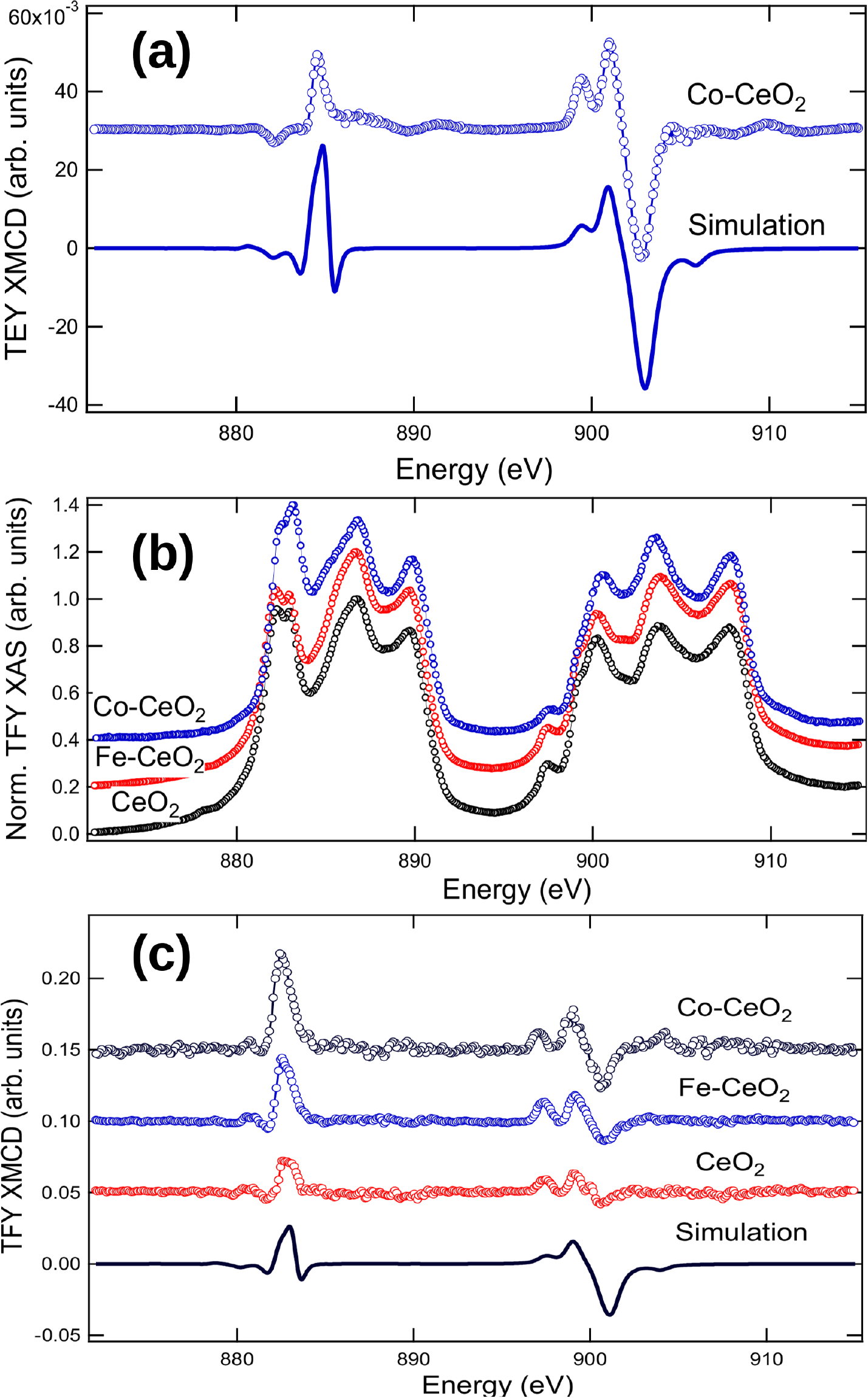}
\caption{Ce $M_{4,5}$ TEY(surface) and TFY(bulk) XMCD spectra evidencing the magnetic moment at Ce 4$f$ sites. (a) Co-CeO$_2$ TEY XMCD and simulation are shown. (b,c) A comparison of Ce $M_{4,5}$ XAS and XMCD is shown. The origin of the energy axis of the simulated spectra has been chosen to align with the maximum intensity peak of the $M_5$ edge and the simulated spectra is reduced by a factor of twelve to match the experimental intensity of nanoceria. \label{fig:Magnetism_Ce}} 
\end{center}

\end{figure}

Covalent orbitals play a major role in understanding the origin of magnetism. In trivalent Ce compounds such as CeRh$_3$B$_2$ and CeCuSi the magnetism is due to highly localized 4$f$ electrons. By contrast in tetravalent $\alpha$--cerium compounds CeFe$_2$ or CeCo$_5$, the magnetism is from hybridization between 4$f$ and conduction electrons\cite{Schille.1994}. The results of density functional theory calculations (LDA+U, LSDA+U, LDA/GGA + U) of nanoceria are controversial. Some studies support charge localization in the oxygen vacancies\cite{Ge.2008,Han.2009} as the source of the magnetism. Other studies identify Ce vacanceis\cite{Fernandes.2009,Zhan.2012} as responsible for ferromagnetism (via superexchange between localized electrons in vacancies and neighboring Ce sites). Finally, some challenge both arguments\cite{Keating.2009}, leaving the question unresolved. Identifying the origin of magnetism in nanoceria (via bulk magnetization techniques such as magnetometry and susceptometry) is complex due to the challenges in decoupling the contributions from Ce mixed valence states and oxygen vacancies. X-ray magnetic circular dichroism (XMCD) is a sensitive tool to investigate the source of magnetism at an elemental atomic level via the excitation of core level electrons to unoccupied states above the Fermi level ($E_F$). XMCD experiments have the advantage of being site and orbital selective due to the electric (or quadrupole) selection rules. To gain insights into the role of the 4$f$ electrons' contribution (conduction or hybridized) we performed surface and bulk sensitive XMCD measurements simultaneously using total electron yield (TEY) and total fluorescence yield (TFY) over the $M_{4,5}$ edges; TEY probes the first $\sim$2~nm of the surface while TFY measures the complete sample but is prone to self-absorption effects\cite{fuhrman2018diamagnetic}.  In Fig.~\ref{fig:Magnetism_Ce}a we present the 10~K artifact free\cite{artifactfree} $\pm$5~T XMCD spectra, the TEY Co-CeO$_2$ Ce $M_{5,4}$ XMCD, is most representative due to the least amount of surface charging.  Note that ceria is a poor conductor, and Co-CeO$_2$'s conductivity is high compared to that of Fe-CeO$_2$ and CeO$_2$ which made it difficult to measure a clean XMCD spectra in TEY for the Fe and CeO$_2$ samples. Both TEY and TFY XMCD spectra clearly identify that the Ce 4$f$ electrons unambiguously carry a magnetic moment on both the surface and in the bulk.

\begin{table}[b!]

\begin{tabular}{c|c|c|c|c}
	                     &  Co-CeO$_2$ & CeO$_2$  	&  Fe-CeO$_2$   &  Co-CeO$_2$   \\ 
                         &  TEY/surface                & TFY               & TFY                  &  TFY           \\ \hline
    $\braket{L_z} (\hbar)$ &   -0.24(1) & -0.24(1)    & -0.36(2)  &  -0.48(2)        \\ \hline
    $\braket{S_z} (\hbar)$ &   0.03(1)  & 0.03(1)     &  0.05(1)  &   0.06(2)        \\ \hline
	$\braket{J_z} (\hbar)$ &   -0.21(1) & -0.21(1)    & -0.32(2)   & -0.42(2)        \\ \hline
    $\braket{L_z}/\braket{S_z}$ & -8 & -8   & -7.2  & -8 
\end{tabular}
\caption{Contributions of the z-component of the orbital and spin magnetic moments obtained from the TEY (surface) XMCD simulations of Co-CeO$_2$ and TFY (bulk) XMCD of CeO$_2$, Fe-CeO$_2$ and Co-CeO$_2$ nanocrystallites.\label{tab:XMCD_tab}}
{
   }
\end{table}

To quantify the magnetic moment, XMCD spectra were simulated using Xclaim\cite{Fernandez.2015} for the 3$d^{10}$4$f^1$ $\rightarrow$ 3$d^{9}$4$f^2$ transition in the atomic limit. The contributions of the XMCD spectral orbital and spin magnetic moments obtained from the surface and bulk contributions are in table~\ref{tab:XMCD_tab}. This dichroic signal is explicit evidence of Ce sites carrying magnetizable moments. In general, the spectral shape of the Ce $M_{4,5}$ edges are indicative of a ground state total angular momentum ($J = \frac{5}{2}$ for a pure state 4$f^1$ state). Any changes in the XMCD spectral shape can be attributed to different values of $J$ contributing to the ground state\cite{Schille.1994,Van.1986}. It is important to note that the simulated spectra are for a pure $J = \frac{5}{2}$ state are not in complete agreement with experiment (e.g.~Fig.~\ref{fig:Magnetism_Ce}a -- negative peak present at the M$_5$ edge and an overestimation(underestimation) of M$_5$(M$_4$) dichroic signals). Interestingly, nanoceria's Ce $M_{4,5}$ XMCD spectral line shape is different from CeRh$_3$B$_2$ and CeCuSi\cite{Schille.1994} (where the ground state is pure $J = \frac{5}{2}$ and magnetism is due to 4$f$ conduction electrons) but quite similar to the XMCD spectra of CeFe$_2$ and a Ce/Fe multilayer (ground state is a mixture of $J = \frac{5}{2}$ and $J = \frac{7}{2}$\cite{Schille.1994,Finazzi.1995}). This is indicative of Ce 4$f$ electrons being strongly hybridized with the O 2$p$ valence band in a mixed ground state of $J = \frac{5}{2}$ and $J = \frac{7}{2}$. At the $M$ edges, although the TFY XAS signal is distorted\cite{Butorin.1995} because of self absorption (Fig. \ref{fig:Magnetism_Ce}b) the TFY XMCD (Fig. \ref{fig:Magnetism_Ce}c) signal is similar to TEY XMCD (surface). The TFY XMCD magnitude increases in the order of CeO$_2$~$<$~Fe-CeO$_2$~$<$~Co-CeO$_2$. Results identify that foreign ions with intrinsic moments (such as Fe and Co) enhances (see SM) the overall magnetic moment at Ce 4$f$ increases (Table.~\ref{tab:XMCD_tab}).

XMCD measurements (atomic magnetism) identify the average magnetic moment as 0.18~$\mu_B$/Ce\cite{Moment}, and if all Ce 4$f$ magnetic states are contributing to the ferromagnetism, the $\sim$20~nm CeO$_2$ crystallites are expected to show $\sim$2000~$\mu_B$/crystallite. In contrast SQUID magnetometry measures the magnetization from the Ce 4$f$, $V_{orb}$, and hybridization contributions with $M_s$=2~$\mu_B$/crystallite identifying that the ferromagnetic volume fraction is only 0.1\%(see SM for XMCD and SQUID magnetometry moment calculations). Clearly, not all Ce 4$f$ states are involved in the magnetism; only the fraction associated with the $V_{orb}$ and/or hybridization are responsible. It follows that because of the low fraction, only Ce 4$f$ -- O 2$p$ states that are captured in the delocalized $V_{orb}$ are associated.

\begin{figure}[t!]

\begin{center}		
\includegraphics[scale=0.45]{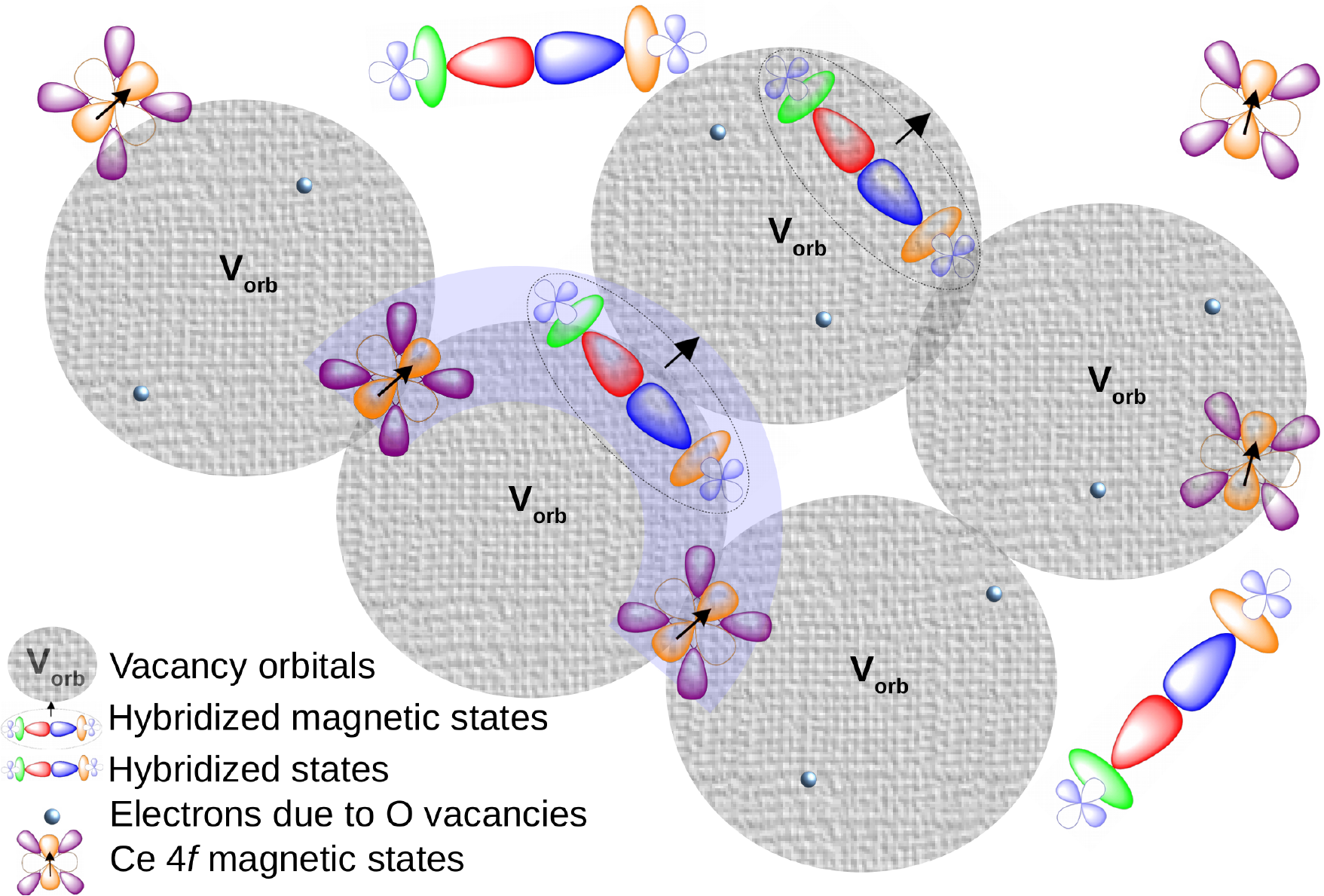}
\caption{Graphical illustration of the magnetic model. Ce 4$f$ magnetic states, $V_{orb}$, and hybridized Ce 4$f$ and O~$2p$ states are shown. Ce 4$f$ states captured in the $V_{orb}$ polarizes the hybridized states and provides a channel to mediate the ferromagnetism. Shaded region illustrates the magnetic exchange process as discussed in the text. \label{fig:Hybridization}} 
\end{center}

\end{figure}

The radial extent of Ce 4$f$ orbitals\cite{Barandiaran.2003} are very small (0.54~\AA) and that limits the Ce 4$f$ -- O 2$p$ covalent mixing to be relatively low as supported by various DFT/LDA/GGA calcuations\cite{Paier.2013}. However, the size (0.5~to~0.8~nm diameter) of the $V_{orb}$ are large (see SM for calculation) and less localized compared to the Ce 4$f$ states. This is consistent with first principle calculations that found the size of $V_{orb}$ at $\sim$1.0~nm\cite{Herng.2010}. Note that only the trapped Ce 4$f$ states in the $V_{orb}$ can polarize spin moments (due to their delocalized nature) on the hybridized states and be responsible for the long range ferromagnetic order. The residual 4$f$ states that are not in the vicinity of $V_{orb}$ cannot contribute to the ferromagnetism due to the lack of the hybridized magnetic states. If the number of $V_{orb}$ are constant, introducing foreign transition metal ions (Fe or Co) impacts Ce 4$f$ -- O 2$p$ hybridization and further promotes a robust, yet weak, ferromagnetism. Figure \ref{fig:Hybridization} shows the illustration of this microscopic model. This description is consistent with the observation that air or O$_2$ annealed d$^0$ nanoscale magnetic oxides exhibit reduced or annihilated magnetism\cite{Sundaresan.2006,Hong.2006,Ackland.2018}, as O$_2$ fills the vacancies resulting in a deficiency of $V_{orb}$ coupling channels. 

In summary, we have found a possible pathway to explain the origin of ferromagnetism in the dilute magnetic oxide nanoceria. Using a combination of electronic structure, elemental and bulk sensitive magnetism techniques we show that $V_{orb}$, Ce 4$f$ spin and orbital angular momentum, and hybridization with O~$2p$ states are crucial for the magnetic ordering. The concept of magnetism from hybridized Ce 4$f$ -- O 2$p$ states in trapped $V_{orb}$ is a missing link to understand the ferromagnetism in nanoceria. In closing, this work provides unambiguous experimental evidence of the origin of ferromagnetism in nanoceria, and demonstrates that this hybridization concept may be a solid foundation from which to explain the unexpected ferromagnetism in ZnO, HfO$_2$, Al$_2$O$_3$, In$_2$O$_3$, SnO$_2$ and many other dilute magnetic oxides and semiconductors (where O 2$p$ hole states are key players, and their hybridization with host or guest metal ions changes the density of states) that present similar magnetism.

\acknowledgments
We thank Dr.~D. J. Keavney for assistance with the XAS and XMCD measurements, and Dr.~S. M. Heald for helping with the XAFS measurements on Co-CeO$_2$. VKP and JvL acknowledge funding from the Natural Sciences and Engineering Research Council of Canada (RGPIN-2018-05012) and the Canada Foundation for Innovation. Use of the Advanced Photon Source at Argonne National Laboratories was supported by the US DOE under contract DE-AC02-06CH11357. 


\bibliography{Ceria_Arxiv.bib}

\begin{thebibliography}{47}
\bibitem [{art()}]{artifactfree}%
  \BibitemOpen
  \href@noop {} {}\bibinfo {note} {The XMCD spectra were obtained as the
  difference of the corrected x-ray-absorption spectra for both left and right
  circularly polarizations. To make sure that the final XMCD spectra were free
  of experimental artifacts, such as those from charging effects, spectra were
  also collected for the opposite applied field direction.}\BibitemShut {Stop}%
\end{thebibliography}


\end{document}



\pagenumbering{gobble}

\clearpage

\pagenumbering{arabic}

\noindent
\textbf{X-ray diffraction patterns of CeO$_2$, Fe-CeO$_2$ and Co-CeO$_2$}

X-ray powder diffraction patterns were collected using a Bruker D8 Discover with Cu $K_{\alpha}$ radiation in a Bragg-Brentano geometry under ambient conditions. The diffraction patterns were collected on dried nanoparticle samples on a zero-background quartz slide using a rotation stage. Lattice parameters and the Scherrer broadening effects were determined using Rietveld refinement with Fullprof\footnote{Rodríguez-Carvajal, Juan. Physica B: Condensed Matter 192, no. 1-2 (1993): 55-69}.

\begin{figure}[h!]

\begin{center}		
\includegraphics[scale=0.45]{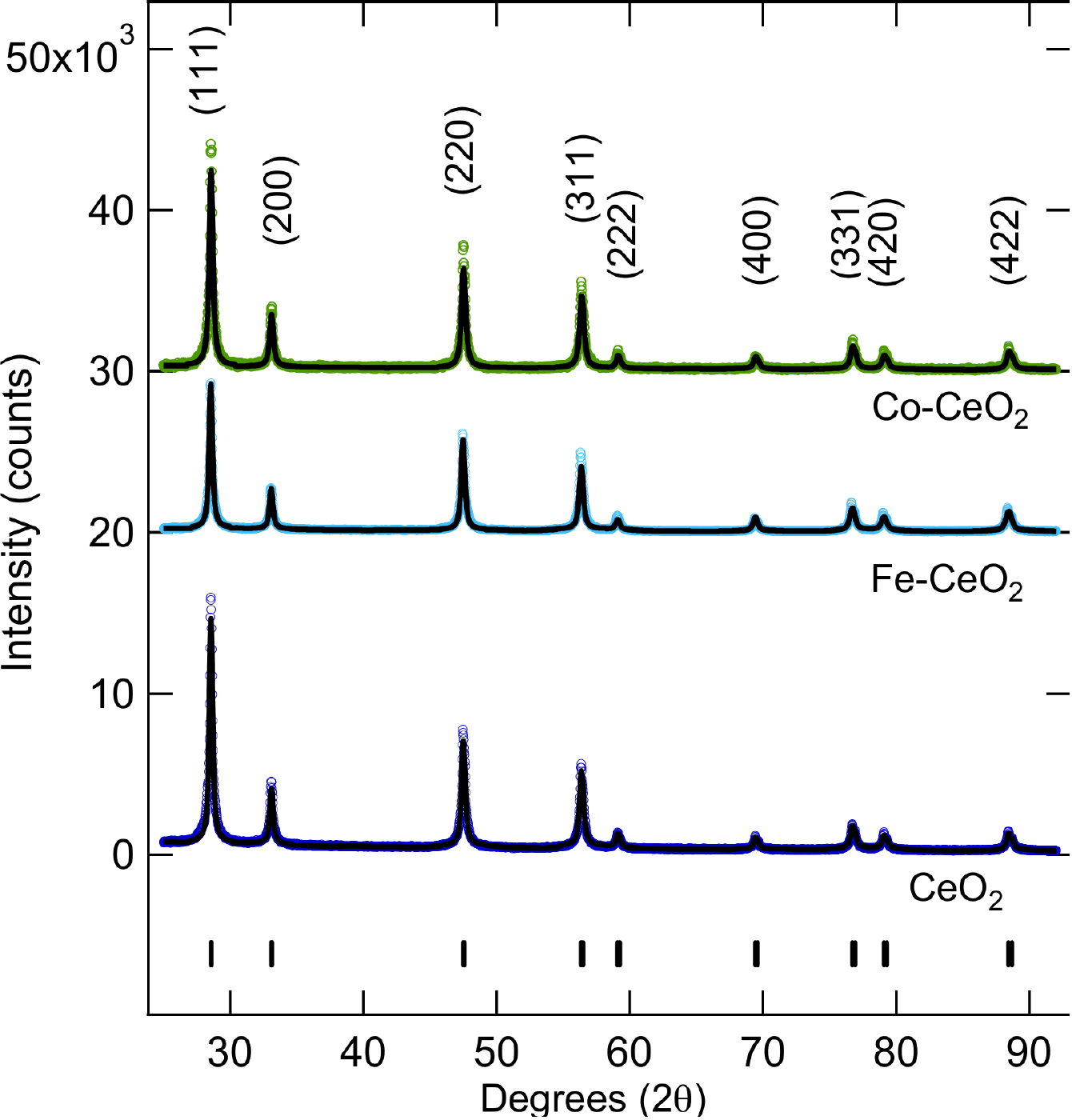}
\includegraphics[scale=0.425]{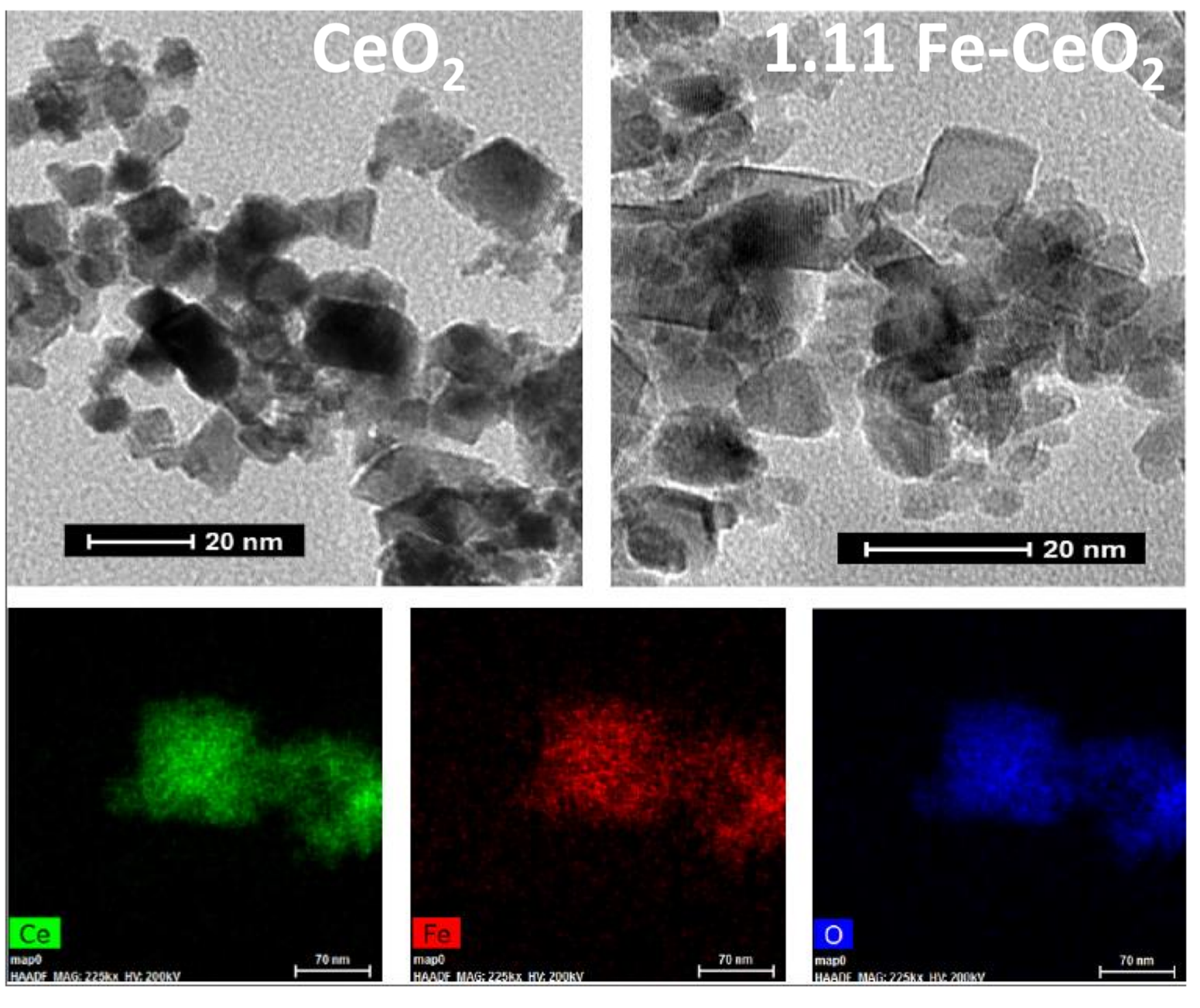}
\caption*{Figure S1: (Left) Room temperature XRD patterns of nanoceria, Fe-CeO$_2$ and Co-CeO$_2$ nanocrystals. The (hkl) indices of the structure are labeled. Refinement of nanoceria, Fe-CeO$_2$ and Co-CeO$_2$ crystallites (solid lines) are presented; the short vertical bars indicate the position of Bragg reflections used in the refinement. (Right) Transmission electron micrographs and high-angle annular dark-field images of nanoceria and Fe/nm$^2$ nanoceria are presented.\label{fig:XRD}}
\end{center}
\end{figure}

\begin{table*}[h!]

\caption*{Table S1: Crystalline (nanoparticle) size (nm) and lattice constant (\AA) from XRD pattern refinements. ICP wt\%'s of Fe and BET measurements for nanoceria and Fe-CeO$_2$ and Co-CeO$_2$ are presented. \label{tab:RR}}
\resizebox{14cm}{!}{
\begin{tabular}{|c|c|c|c|c|c|c|c|}
\hline
	Sample     & size (nm)	& $a$ (\AA, XRD)     & ICP wt\%  & BET (m$^2$/g)	\\ \hline 
    CeO$_2$    & 21.0 $\pm$ 0.3           & 5.412~$\pm$~0.001 &  0        & 55           \\ \hline
    Fe-CeO$_2$ & 23.4 $\pm$ 0.5           & 5.412~$\pm$~0.001 &  0.57     & 55           \\ \hline
    Co-CeO$_2$ & 20.0 $\pm$ 0.4           & 5.410~$\pm$~0.001 &  3.57     & 55           \\ \hline
\end{tabular}
   }
\end{table*}

\clearpage


\pagenumbering{gobble}

\clearpage

\pagenumbering{arabic}

\noindent
\textbf{X-ray absorption fine structure (XAFS) of CeO$_2$, Fe-CeO$_2$ and Co-CeO$_2$}

The XANES and EXAFS measurements of Ce $L_3$ and Fe K-edge measurements were performed at beamline 20-ID-B,C of the Advanced Photon Source located at Argonne National Laboratory and Co K-edge were at beamline 20-BM-B. A Si (111) double crystal monochromator and KB mirrors were used to provide monochromatic microbeam of the size 100 $\mu$m. Measurements on Ce $L_3$ edge are done in transmission geometry. Detectors were ionization chamber based. Calibration was done using Cr foil. 

Due to low Fe and Co contents Fe $K$-edge and Co $K$-edge measurements are done in TFY geometry. A 13 element solid-state detector was used to monitor the fluorescence x-rays. In both transmission and fluorescence geometries specimens were a thin powder prepared using Kapton tape. All spectra were analyzed using the ATHENA and ARTEMIS software programs\footnote{B. Ravel and M. Newville, Journal of synchrotron radiation 12, 537 (2005)}. The theoretical calculation of the phase shifts and backscattering amplitudes for specific atom pairs were obtained using the FEFF program\footnote{J. J. Rehr and R. C. Albers, Reviews of modern physics 72, 621 (2000)} based on the crystallographic data of CeO$_2$. EXAFS of CeO$_2$ was fitted to its crystal structure to obtain the amplitude reduction factor ($S_0^2$). EXAFS analysis at the Ce $L_3$ edge includes only the shell of the Ce -- O single scattering path due to the small energy separation between $L_3$ and $L_2$ edges.   

\begin{table}

\caption{Parameters from the shell fitting of Ce $L_3$ edge EXAFS analysis of CeO$_2$, Fe-CeO$_2$ and Co-CeO$_2$ nanocrystallites. \label{tab:EXAFS}}
{
\begin{tabular}{|c|c|c|c|c|c|c|c|}
\hline
	sample         & $N$	& $E_0$ (eV)     & $\sigma^2_{Ce-O1} (\AA^2)$ & $R_{Ce-O1}$ (\AA)	\\ \hline 
       CeO$_2$     & 5(1)      & 6(1) &  0.001(1)       & 2.32(2)           \\ \hline
   	Fe-CeO$_2$     & 7(2)      & 5(2) &  0.005(4)       & 2.30(4)           \\ \hline
    Co-CeO$_2$     & 8(1)      & 4(1) &  0.007(2)       & 2.31(4)           \\ \hline
\end{tabular}
   }
\end{table}

\textbf{Ce $M_{4,5}$ edge XAS and XMCD of CeO$_2$, Fe-CeO$_2$ and Co-CeO$_2$}

The XAS and XMCD measurements were conducted at beam line 4-ID-C of Advanced Photon Source located at the Argonne National Laboratory. Ce $M_{4,5}$ edge XAS and XMCD measurements were collected simultaneously in total electron yield and total fluorescence yield with circularly polarized x-rays. XMCD spectra were obtained both in left and right circularly polarized x-rays and measurements were taken for both $\pm$ 5T at 10 K and then we take difference of these two XMCD spectra to eliminate polarization dependent systematic errors (artifact free XMCD).

\begin{figure}[h!]

\begin{center}		
\includegraphics[scale=0.5]{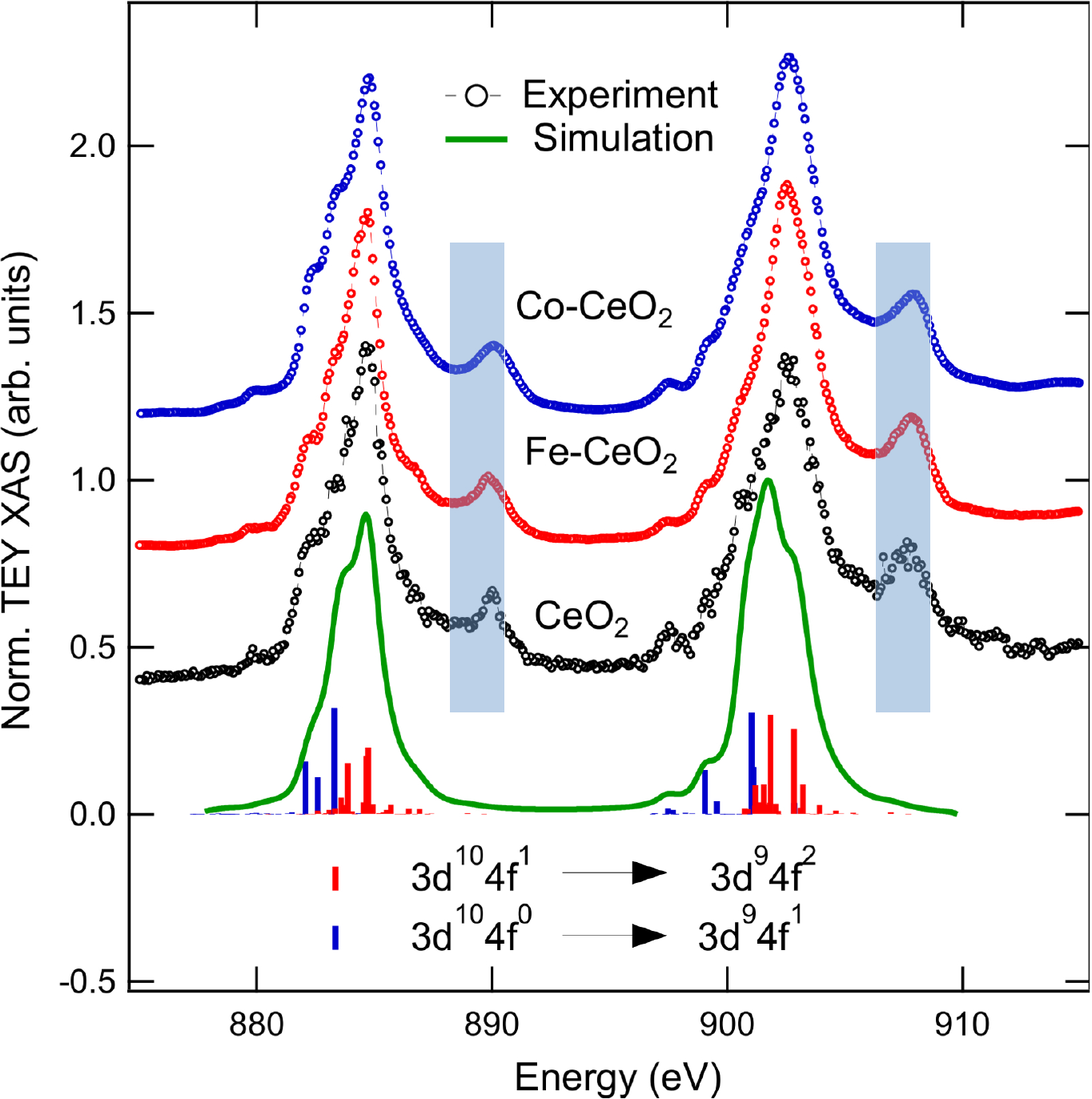}
\caption*{Figure S2: Ce M$_{4,5}$ edge XAS data  and simulation with ionic limit. Higher energy satellite peaks region is shaded to show that the spectra could not be modelled successfully with 3$d^{10}$4$f^0$ (initial) and 3$d^9$4$f^1$ (final) states and charge-transfer effects needs to be included.\label{fig:NoCTM_CeO2}}
\end{center}
\end{figure}

XAS simulations of Ce $M_{4,5}$ edge 3d$^{10}$4f$^1$ $\rightarrow$ 3d$^{9}$4f$^2$ transitions were simulated using CTM4XAS5.5 GUI with Slater integrals $F_{ff}$ at 79\%, $F_{df}$ and $G_{df}$ at 100\% atomic values and 3$d$ spin orbit coupling at 98\%. A Gaussian broadening of 0.20 eV was applied to account for instrumental broadening and Lorentzian broadening of 0.4 eV and 0.2 eV were applied for $M_5$ and $M_4$ edges, respectively. Experimental spectra could not be modeled sucessfully by in atomic(ionic) limit and charge transfer effects needs to considered.

Charge transfer simulations for 3$d^{10}$4$f^0$ + 3$d^{10}\underline{L}$4$f^1$ $\rightarrow$ 3$d^{9}$4$f^1$ + 3$d^{9}\underline{L}$4$f^2$         ($\underline{L}$ is the hole in the O 2$p$ band) were implemented using the commandline (as described in the CTM4XAS manual). For this configuration, $\Delta_{IS}$ was set to 2.0 eV, $\Delta_{FS} = -2.5$ eV, and T$_{IS}$ = T$_{FS}$ = 0.77 eV. Gaussian broadening of 0.20 eV was applied to account for instrumental broadenings and Lorentzian broadenings of 0.50 and 0.60 eV were applied to the $M_5$ and $M_4$ edges, respectively. 

\clearpage

\noindent
\textbf{Ce M$_{4,5}$ edge TEY XMCD of CeO$_2$ and Fe-CeO$_2$}

\begin{figure}

\begin{center}		
\includegraphics[scale=0.65]{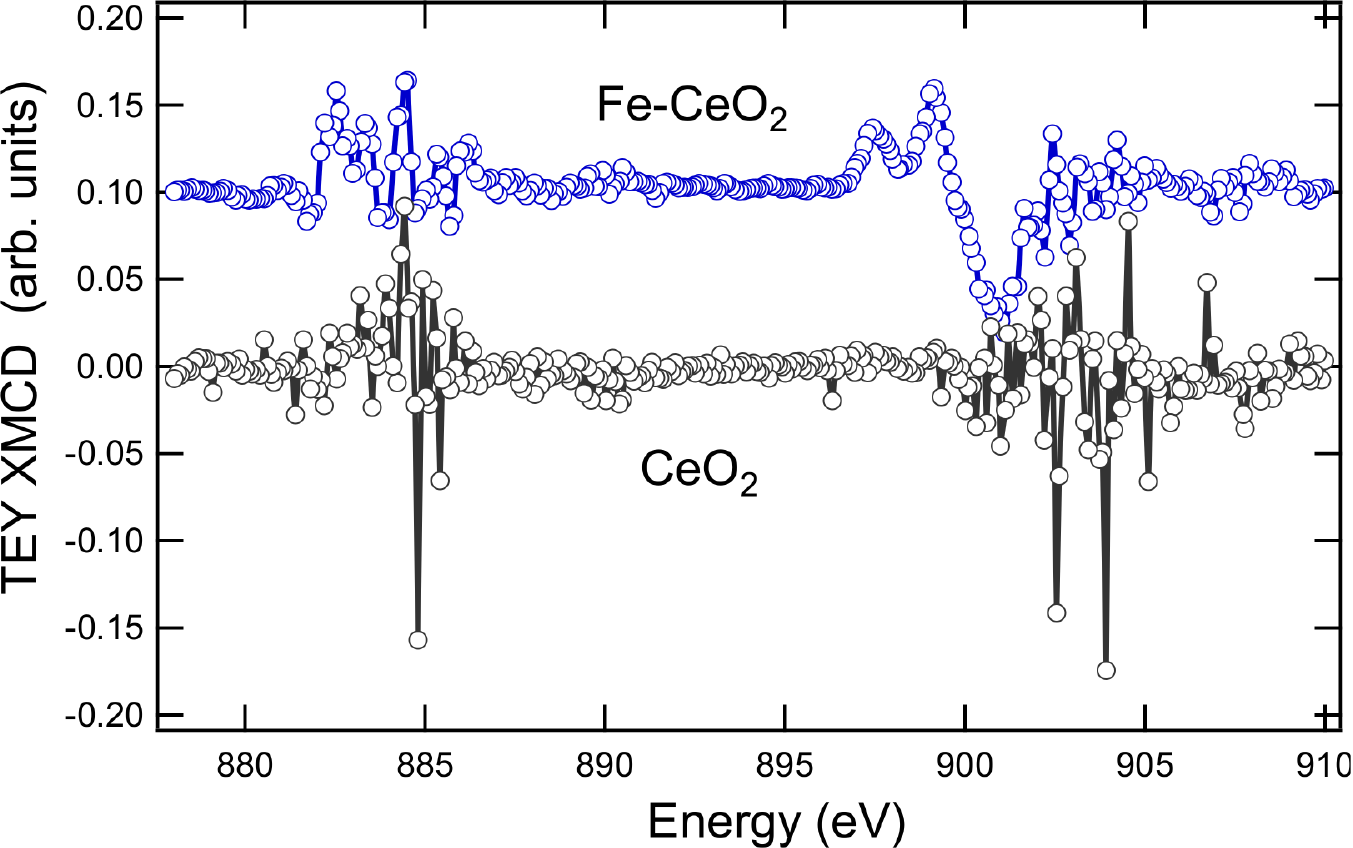}
\caption*{Figure S3: Ce $M_{4,5}$ edge TEY XMCD signal affected by charging effects. \label{fig:Fe_Ledge}}
\end{center}
\end{figure}

\noindent
\textbf{Fe L$_{3,2}$ edge XAS and XMCD of Fe-CeO$_2$}

\begin{figure}

\begin{center}		
\includegraphics[scale=0.6]{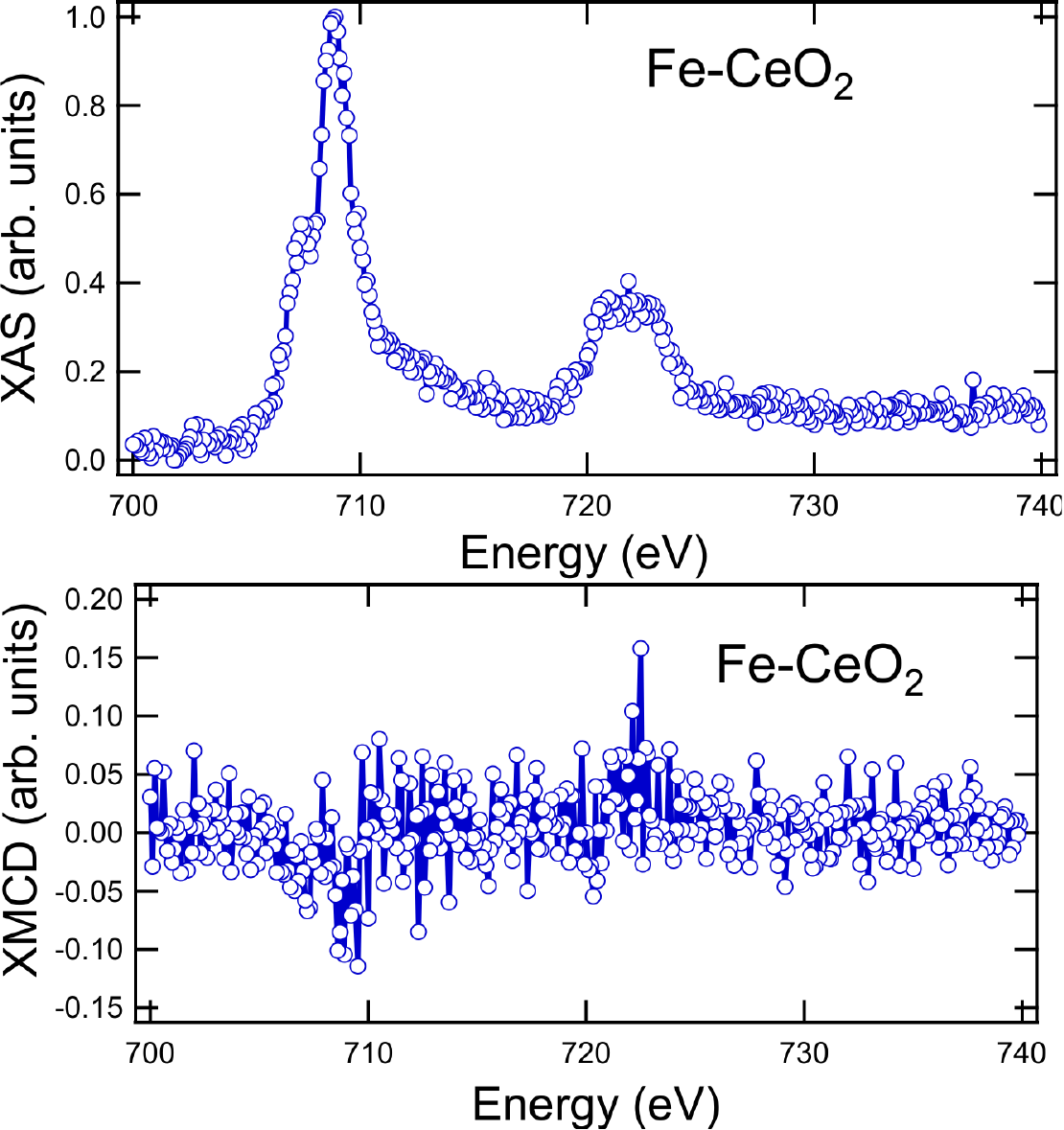}
\caption*{Figure S4: Fe $L_{3,2}$ edge XMCD shows no moment.\label{fig:Fe_Ledge}}
\end{center}
\end{figure}
\clearpage

\noindent
\textbf{Co L$_{3,2}$ edge XAS and XMCD of Co-CeO$_2$}

\begin{figure}

\begin{center}		
\includegraphics[scale=0.6]{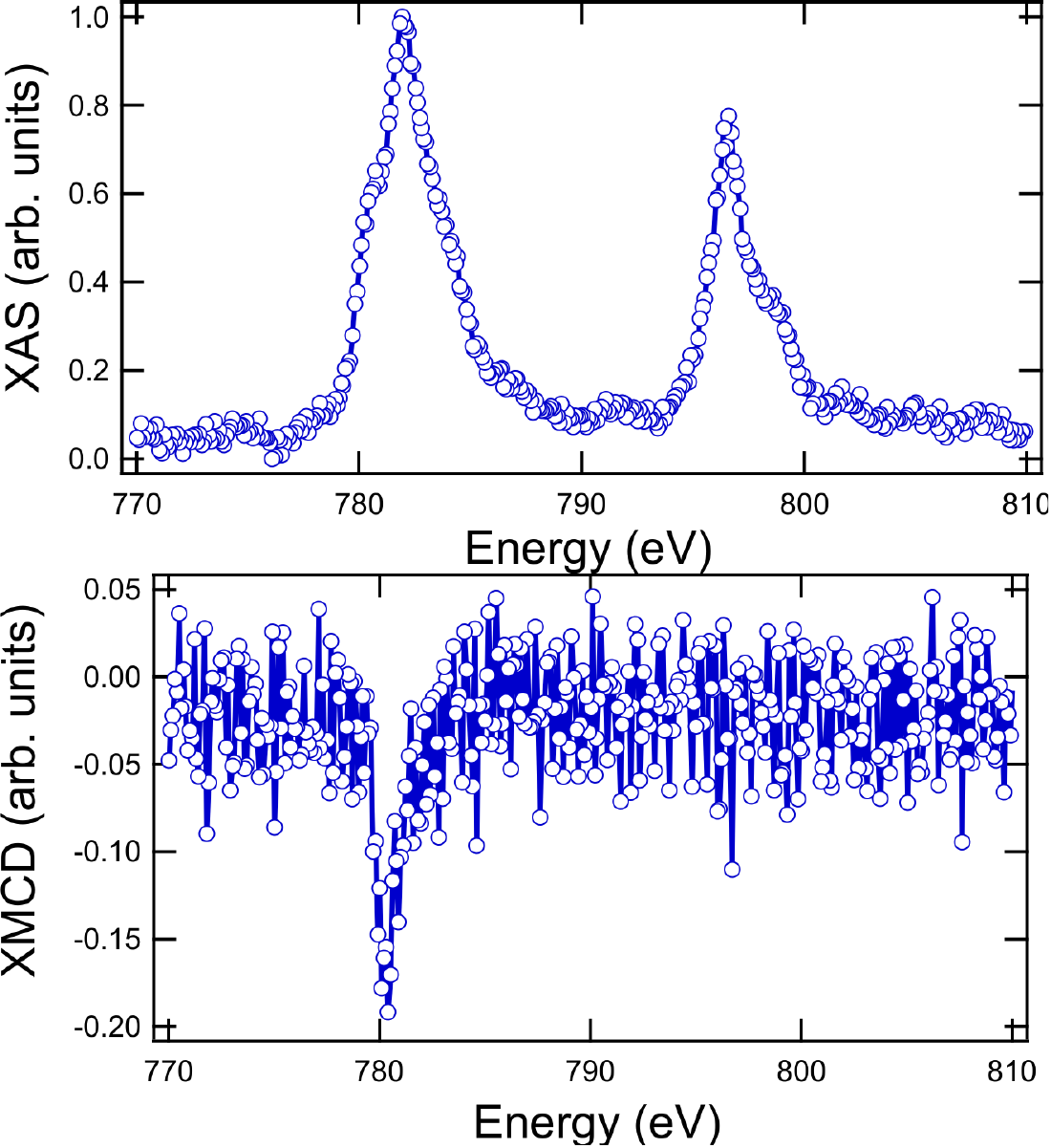}
\caption*{Figure S5: Co $L_{3,2}$ edge XMCD shows moment.\label{fig:Co_Ledge}}
\end{center}
\end{figure}

\textbf{Magnetic measurements.}
Magnetometry experiments were performed using a Quantum Design magnetic properties measurement system (MPMS XL-5 using the Reciprocating Sample Option (RSO)). The samples were mounted in low background NMR (Norell high resolution S-5-20-8) tubes. $M(\mu_0H)$ measurements\footnote{Considering that only a fraction of the volume of nanoparticles are sponteneously ferrromagentic the $M_s$'s presented in terms of volume magnetization (Am$^{-1}$)} were done at 300 K and data are corrected for the high field susceptibility.  

\begin{figure}[h]

\begin{center}		
\includegraphics[scale=0.40]{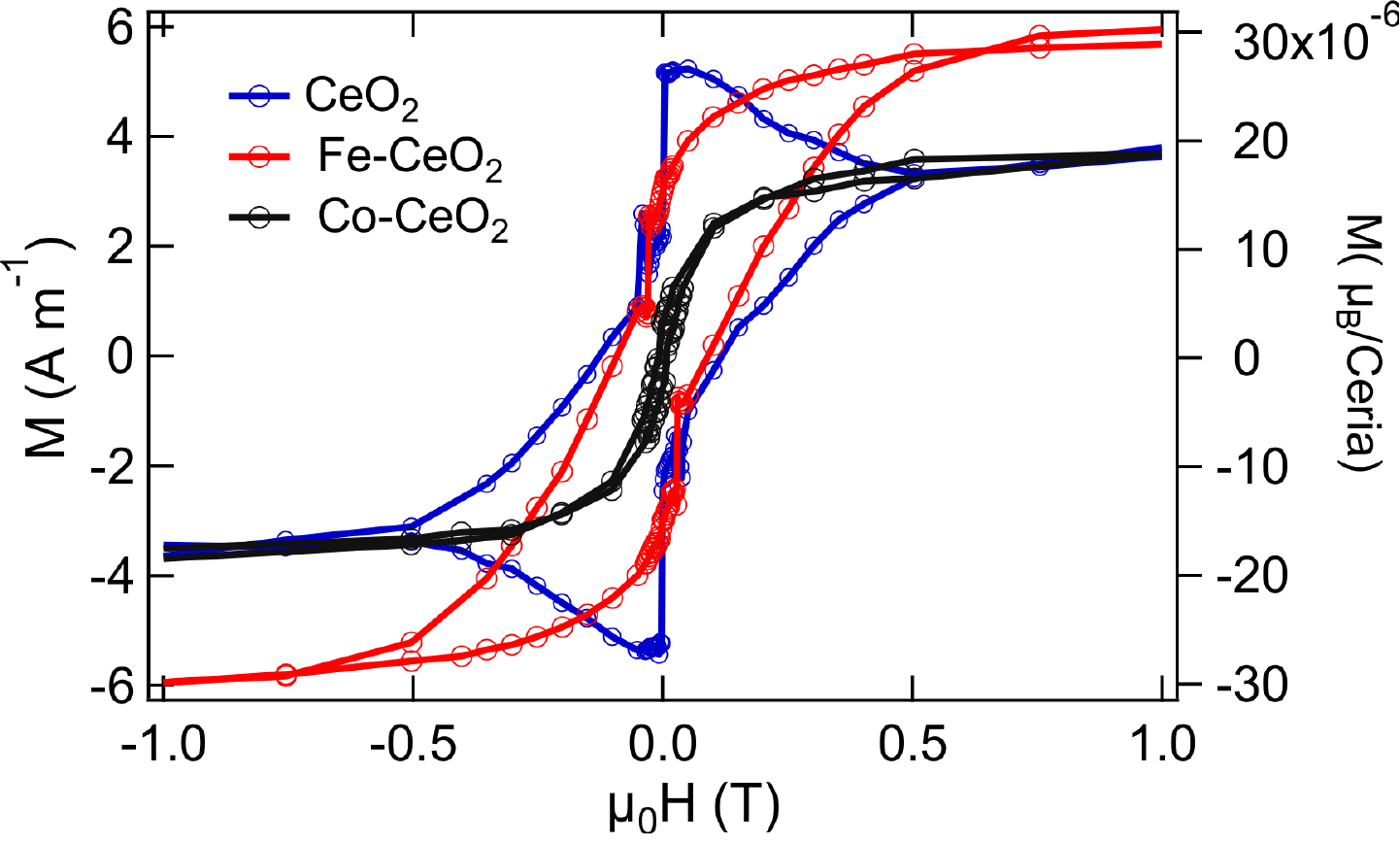}
\caption*{Figure S6: $M(\mu_0H)$ of CeO$_2$, Fe-CeO$_2$ and Co-CeO$_2$ nanocrystallites measured at room temperature. $M(\mu_0H)$ of nanoceria shows a coercivity of $\sim$50~mT and saturation magnetization ($M_s$) of $\sim$4~Am$^{-1}$. Fe-CeO$_2$ and Co-CeO$_2$ nanocrystallites shows coercivity of $\sim$50~mT and $\sim$0~mT and $M_s$' were $\sim$7~Am$^{-1}$ and $\sim$4~Am$^{-1}$.\label{fig:SQUID_Ce}} 
\end{center}

\end{figure}

\clearpage

\noindent
\textbf{M{\"o}ssbauer hyperfine parameters of Fe-CeO$_2$}

Transmission M{\"o}ssbauer spectra were collected at RT using a WissEl constant acceleration spectrometer with a 10-GBq $^{57}$Fe\textbf{Rh} source. The source drive velocity was calibrated using a 6 $\mu$m thick $\alpha$-Fe.

\begin{figure}

\begin{center}		
\includegraphics[scale=0.7]{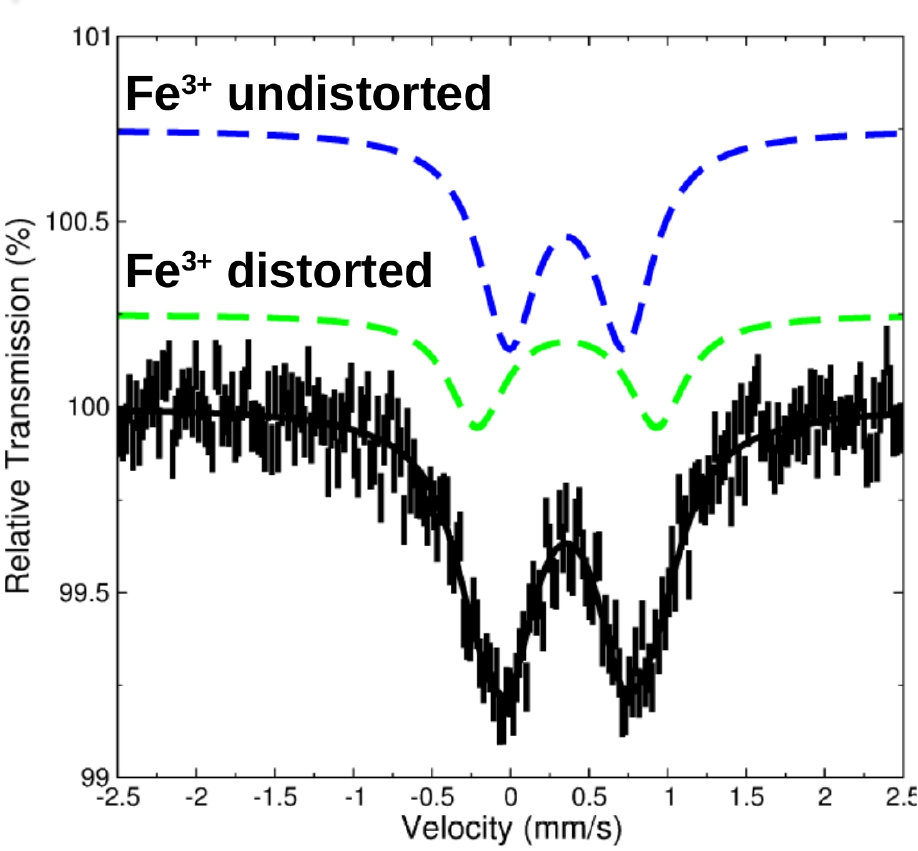}
\caption*{Figure S7: M{\"o}ssbauer spectra of Fe-decorated CeO$_2$ systems. The solid lines represent the fitted spectra (left) and subspectra components (right) used for the undistorted (site-I) and distorted (site-II) are shown. \label{fig:Fe_Mossbauer}}
\end{center}
\end{figure}

\begin{table*}[h!]

\caption*{Table S3: M{\"o}ssbauer hyper fine parameters (line width ($\Gamma$), isomer-shift ($\delta$) and quadrupole splitting ($\Delta$) are presented. \label{tab:Mossy}}
\resizebox{18cm}{!}{
\begin{tabular}{|c|c|c|c|c|c|c|c|c|}
\hline
 Sample   & $\Gamma$ (mm/s)$(\pm 0.02)$ & $\delta$(mm/s) (I) & $\Delta$ (mm/s) (I) & $\delta$(mm/s) (II) & $\Delta$ (mm/s) (II) \\ \hline 
       Fe-CeO$_2$ & 0.22  & 0.36~$\pm$~0.01  &  0.74~$\pm$~0.06 & 0.36~$\pm$~0.01 &  1.14~$\pm$~0.07           \\ \hline
\end{tabular}
   }
\end{table*}


M{\"o}ssbauer spectra results identify that Fe is in +3 oxidation state of undistorted and distorted environments. The absence of hyperfine field ($B_{hf}$) is indicative of no secondary phase formation (consistent with XRD results).

\clearpage

\noindent
\textbf{Fe and Co K-edge EXAFS analysis of Fe-CeO$_2$ and Co-CeO$_2$}

The theoretical calculation of the phase shifts and scattering for specific atom pairs were obtained using the FEFF program. The amplitude reduction factor $S_0^2$ was obtained from $\alpha$-Fe$_2$O$_3$ and Co$_3$O$_4$ for Fe and Co K-edge respectively. EXAFS shell fitting results illustrate that Fe neighbors are O atoms. Similarly Co first neighbors are O atoms, second and third neighbors are Co atoms. The enhancement of magnetic moment at Ce sites increased due to hybridization of O 2$p$ valence band. 

\begin{figure}

\begin{center}		
\includegraphics[scale=0.50]{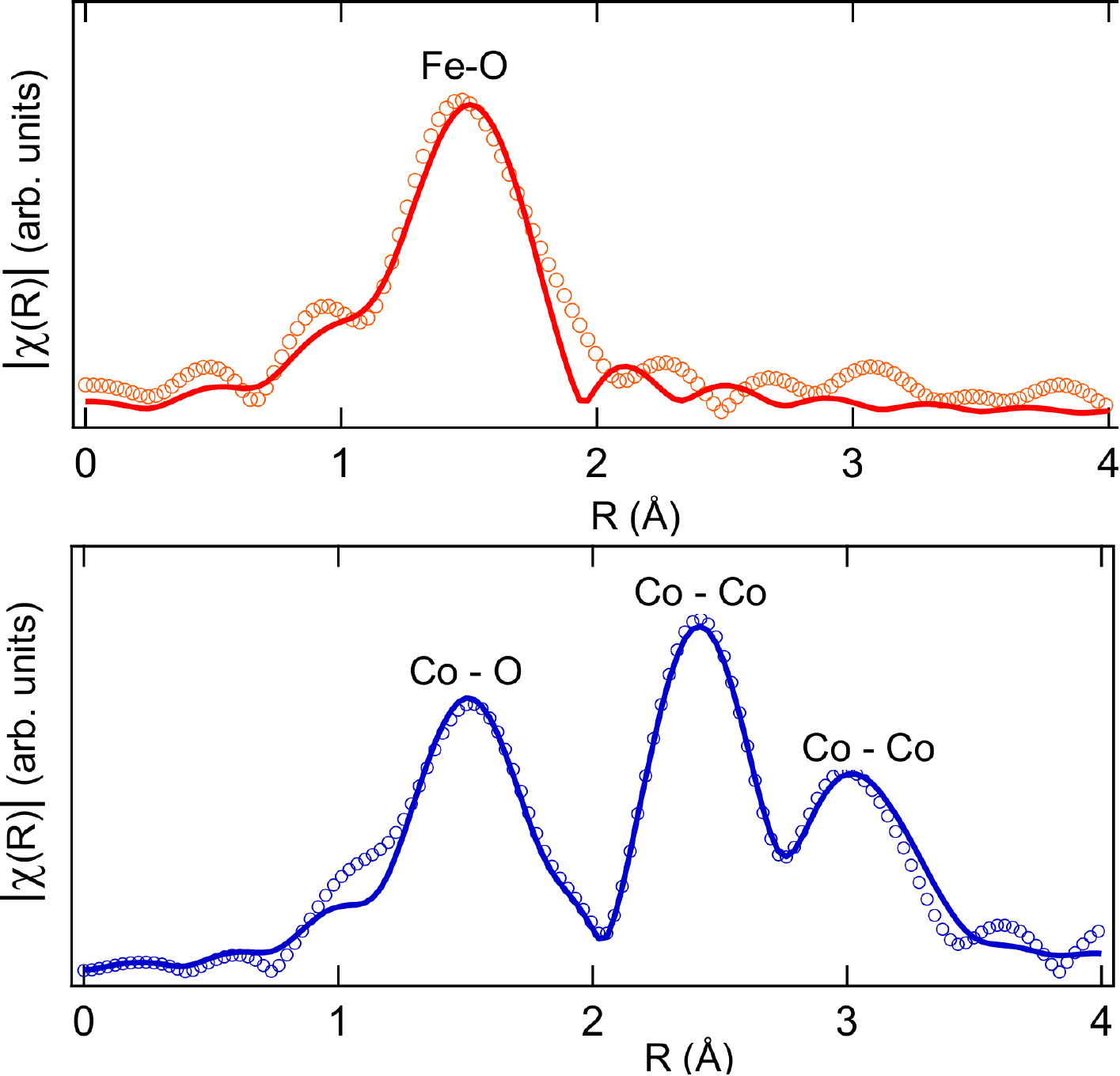}
\caption*{Figure S8: Fourier transformed magnitudes and fits of Fe-CeO$_2$ and Co-CeO$_2$ EXAFS data is shown.\label{fig:SQUID_Ce}} 
\end{center}

\end{figure}

\begin{table}[b!]

\caption{Parameters from the shell fitting of Fe $K$ edge EXAFS analysis of Fe-CeO$_2$. \label{tab:Fe_EXAFS}}
\resizebox{10cm}{!}{
\begin{tabular}{|c|c|c|c|c|c|c|c|}
\hline
	Shell         & $N$	& $E_0$ (eV)     & $\sigma^2 (\AA^2)$ & $R$ (\AA)	\\ \hline 
   	Fe -- O     & 3(1)      & -5(2) &  0.008(2)       & 1.97(2)           \\ \hline
    \end{tabular}
   }
\end{table}
\begin{table}[b!]

\caption{Parameters from the shell fitting of Co $K$ edge EXAFS analysis of Co-CeO$_2$. \label{tab:Co_EXAFS}}
\resizebox{10cm}{!}{
\begin{tabular}{|c|c|c|c|c|c|c|c|}
\hline
	Shell         & $N$	& $E_0$ (eV)     & $\sigma^2 (\AA^2)$ & $R$ (\AA)	\\ \hline 
   	Co -- O     & 8(1)      & -5.7(8) &  0.0020(9)       & 1.925(8)           \\ \hline
    Co -- Co    & 8(1)      & -5.7(8) &  0.0049(9)       & 2.867(1)           \\ \hline
    Co -- Co    & 19(2)     & -5.7(8) &  0.0049(9)       & 3.35(2)           \\ \hline    
\end{tabular}
   }
\end{table}

\clearpage

\noindent
\textbf{Estimating $\mu_B$/crystallite from XMCD and SQUID magnetometry}

The average magnetic moment identified from XMCD is 0.18 $\mu_B$/Ce \footnote{Total moment=$\braket{m^z_{tot}}$=$\frac{-\mu_B}{\hbar}$ ($g_s\braket{S_z}+g_l\braket{L_z}$; $g$-factor $g_s$=2 with the spin moment and $g_l$=1 with the orbital moment [H.C. Siegmann, J. St\"ohr Magnetism. From Fundamentals to Nanoscale Dynamics Springer (2006)].)} (see table 1 in main text for $\braket{L_z}$ and $\braket{S_z}$). A $\sim$20 nm ceria crystallite has a volume of 4.2 $\times$ 10$^{-24}$ m$^3$. From density\footnote{C.I. Kasei Co., Ltd. NanoTek powder.} and molecular weight this consists of $\sim$10$^5$ Ce atoms ($N~=~\frac{N_A \times \rho_{CeO_2} \times V}{M_{CeO_2}}$; where $N_A$=6.023 $\times$ 10$^{23}$ Ce atoms/mol, $\rho_{CeO_2}$=7.1 g/cm$^3$ and $M_{CeO_2}$=172.12 g/mol). If one assumes all Ce 4$f$ states are responsible for the magnetism, that places an upper bound on the moment at $\sim$2 $\times$ 10$^4$ $\mu_B$/crystallite. In nanoceria, each Ce atom can donate four electrons to bonding orbitals with two O atoms. When an oxygen vacancy ($V_O$) is formed, the two electrons previously occupying $p$ orbitals of the O atom are free to distribute. The localized electrons around Ce atoms changes the oxidation state from Ce$^{4+}$ to Ce$^{3+}$ (charge neutralization expression can be written as 2Ce$^{4+}$ + $V_O^{2-}$ $\leftrightarrow$ 2Ce$^{3+}$). This suggests that each $V_O$ is responsible for the creation of two Ce$^{3+}$ ions. Ce $L_3$ edge XANES analysis quantify (see Fig. 1a of main text) the concentrations of Ce$^{3+}$ and $V_O$ as $\sim$20\% and $\sim$10\% respectively. Based on these results, even if one considers only the Ce 4$f$ sites that are neighbouring $V_O$s' are responsible for the ferromagnetism, that puts a lower bound on the moment at $\sim$2 $\times$ 10$^3$ $\mu_B$/crystallite. SQUID magnetometry provides a measure of all magnetic componenets (i.e. not site or element specific). The saturation magnetization of nanoceria is 4 A/m. This magnetization is equivalent to $\sim$6~$\times$~10$^{16}$~$\mu_B$/g \footnote{1 Am$^{-1}$=(10$^{-3}$emu/cm$^3$)/(7.1~g/cm$^3$~$\times$~0.9274$\times$10$^{-20}$emu/$\mu_B$)} which corresponds to each ceria crystallite having a moment of $\sim$2~$\mu_B$. Combinedly XMCD and SQUID magnetometry results show that the ferromagnetic volume fraction is between 0.1 -- 0.01\%.

\clearpage

\noindent
\textbf{Estimating the size of the vacancy orbitals ($V_{orb}$)}

Given the concentration of $V_{orb}$ from Ce $L_3$ edge XANES, the radius of $V_{orb}$ can be estimated\footnote{Herng et al., Phys. Rev. Lett., 105, 207201 (2010)}. In CeO$_2$ each Ce ion has eight surrounding oxygen ions, and each oxygen has four neighboring Ce ions (Fig. S9). We approximate the Ce-lattice as simple face centered cubic (Fig. S9) with 4 Ce's per unit cell (8$\times$1/8 + 6$\times$1/2) and a lattice constant $a$. Similarly, we describe the $V_{orb}$ lattice as simple cubic (Fig. S9) with 1 $V_{orb}$ per unit cell (4$\times$1/4) with a lattice constant $A$. 

\vskip24pt

The $V_{orb}$ concentration is $x=\frac{N_{V_{orb}}}{N_{Ce}}$, where $N_{V_{orb}}$ (=1/$A^3$)  and $N_{Ce}$ (=4/$a^3$) are the density of $V_{orb}$ and Ce respectively, the relationship between lattice constants is $A=a\times(\frac{1}{4x})^\frac{1}{3}$.

\begin{figure}

\begin{center}		
\includegraphics[scale=0.65]{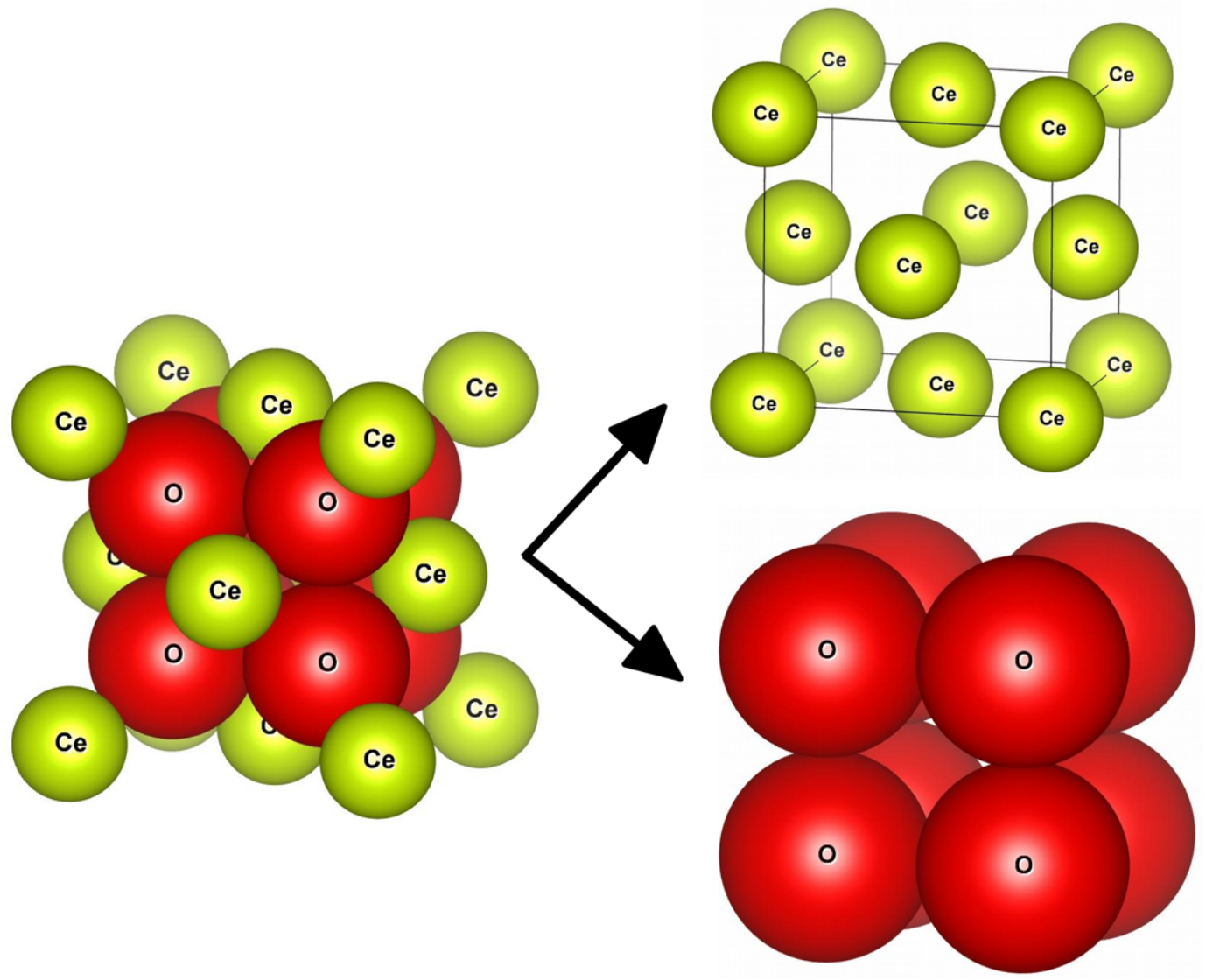}
\caption*{Figure S9: CeO$_2$ crystal structure is shown in space filling style. Ce-lattice is FCC and $V_{orb}$-lattice is simple cubic. Each Ce-lattice consists of 4 Ce's per unit cell and $V_{orb}$-lattice consists of 1 $V_{orb}$ per unit cell.\label{fig:Ceria}} 
\end{center}

\end{figure}
\clearpage

With the relation between $x$, $a$, and $A$, we can estimate the upper and lower bounds of the $V_{orb}$ radius by considering two simple cases, as shown in Fig. S10. For $V_{orb}$ lower bound ($V_{l.b}$) we take radius as the distance at which the orbitals just touch each other. This results in $V_{l.b}=(1/\sqrt{2}) \times(\frac{1}{4x})^\frac{1}{3}a$. For the upper bound we take the radius as the distance at which there is no void, and $V_{u.b}=(1/2) \times(\frac{1}{4x})^\frac{1}{3}a$. Ce $L_3$ edge XANES analysis gives the vacancy fraction at $x=0.1$ (2Ce$^{4+}$ + $V_O^{2-}$ $\leftrightarrow$ 2Ce$^{3+}$) and x-ray diffraction refinements identify the lattice constant, $a \sim 5.412\AA$ (see table. S1), so $V_{orb}$ is between $\sim 0.5$ nm and $\sim 0.8$ nm in diameter.   

\begin{figure}

\begin{center}		
\includegraphics[scale=0.6]{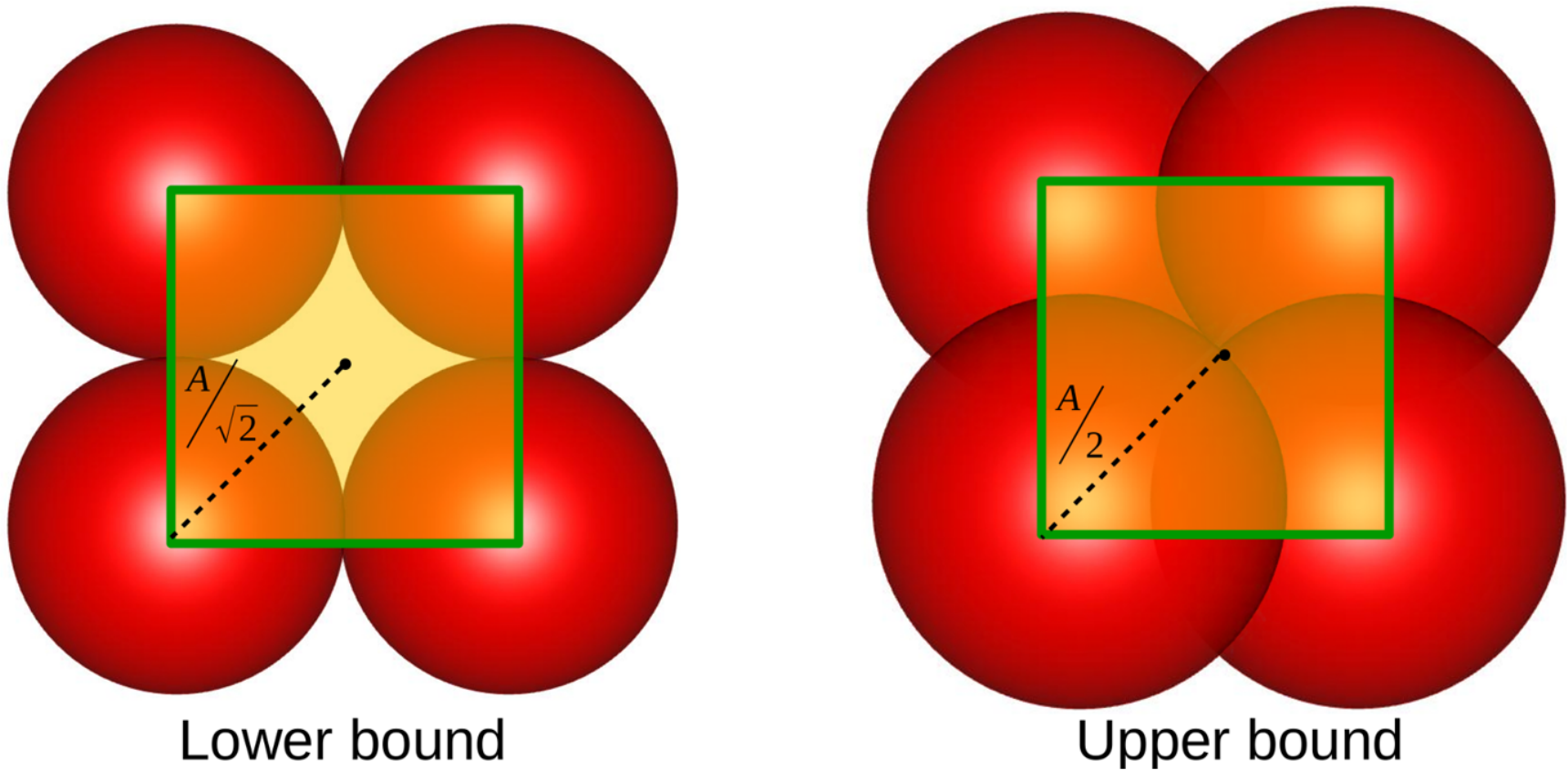}
\caption*{Figure S10: The lower and upper bounds of $V_{orb}$ are shown. In the lower bound the radius of $V_{orb}$ is defined as the distance at which neighboring atoms. Upper bound is the distance at which the $V_{orb}$ fills all space.\label{fig:Vorb}} 
\end{center}

\end{figure}